\documentstyle[12pt]{article}
\begin{document}
\title{Rapidity gap signals in Higgs production at the SSC}
\author{R.S. Fletcher \\ {\small \em Bartol Research Institute,

University of Delaware,Newark DE 19716}\\ T. Stelzer \\ {\small
 \em Physics Dept.

University of Wisconsin-Madison,Madison WI}}
\date{}
\maketitle
\begin{abstract}

We examine the structure of the  underlying event in neutral Higgs
production at the Superconducting-Supercollider (SSC).
Gaps, regions of rapidity containing

no soft particle production, can provide a clean

signature for $W$ boson fusion  to the heavy Higgs. We first

examine the physical
basis of gap production and estimate the  survival probability of
gaps
in the minijet model.
Then, using PYTHIA, and HERWIG we compare gap events to

$W$ pair production from top decay and  $q\bar{q}$ fusion.

We find that, if experimental problems can be overcome, gaps

should provide a small, but clean, signal for

heavy Higgs production at the SSC.
\end{abstract}

\section{Introduction}
One of the  main goals of the Superconducting-Supercollider is

to search for the Higgs particle.  The Higgs is expected to have a
mass in
the range from 100 GeV to at most a few TeV.  Much study has gone
into finding
ways of identifying the Higgs particle  through it's various decay
channels\cite{hhg}.
It has been suggested that the structure of the underlying event
in Higgs production will differ from normal events,
allowing experiments to search for the Higgs by looking for events
where

a `gap' is produced in the central rapidity
region\cite{Doks92,bj92}.

There are three components to particle production in high energy

hadronic interactions. There are high $p_t$

jets caused by the fragmentation of partons

which have undergone a hard scattering. There is an 'underlying
event'
of soft particles which typically  has the structure of a minimum
bias event,

and there are  particles generated by the fragmentation of the soft
partons  radiated from the hard scattering. Although this separation
is
certainly

not  unique, it is useful to discuss these components separately when

considering the production of rapidity gaps in hadronic interactions.

We have used the minijet model to estimate the survival probability

of gaps\cite{bj92},
 and find that around 3\% of events are free of secondary
interactions which would destroy the gap signal.
Next we discuss the properties of the underlying event, and of parton
radiation
in gap events and use
the PYTHIA  and HERWIG
Monte-Carlos\cite{beng87,sjos87b,herwig,marc88} to investigate the
production of gaps
in a full simulation.

\section{Gap production and tagging jets}

There are two main, parton level, processes for producing Higgs
particles

at high energy hadron colliders. First, there is gluon fusion as
shown in
Fig. 1a. Here, the Higgs is coupled to a pair of gluons through a
fermion
loop. Because the Higgs couples to fermions according to their mass,
the
term represented by a top quark in the loop dominates the cross
section.
The cross section  grows with the top quark mass because of this
coupling,
and is large for light higgs particles, because of the large flux of

gluons in the proton. This process dominates Higgs production at the

SSC for high top masses, or Higgs masses less than around a half a
TeV.
 Second, there is the W-boson fusion

reaction, shown in Fig. 1b. Like resonance production in two-photon

interactions at $e^+e^-$ colliders, this process can

be considered in the effective
boson approximation.  In this approximation, incoming quarks

radiate virtual bosons, here W's.  These bosons then

collide, and
annihilate into a Higgs particle.

  At the parton level the

main property of this process is that the outgoing quarks typically
have

significant $p_t$ and  large energy, and therefore

produce jets at large rapidities.  This has led to the suggestion of
using jet
tagging to identify the production of the Higgs\cite{cahn87,barg91}.

There is a difference in the color flow in the diagrams shown

in Fig. 1.  This difference leads to a difference in the underlying
event structure in the  processes. The $W$ fusion diagram can lead to
the

production of a `rapidity gap'.
Gap production occurs when 1) the jets are widely separated, and
2) there is {\em no} particle production in the region between the
jets.
In a typical  high energy event, the region between the jets would be
filled
with  soft  particles from the
underlying event, and by particles  connected with radiation from the

hard scattering that produces the jets. In gap production neither

component exists,  and no particles are produced in the rapidity
region

between the jets.  First we will use a string model for the
underlying event

to describe how particle production is suppressed in the gap, and
then look

at parton radiation in the hard scattering.

\section{Gap production and the underlying  event}

To understand how the underlying event is modeled in a string
Monte-Carlo,
such as PYTHIA
it is easiest to first look at minimum bias events. This is described
in

detail in Ref. \cite{cape81,sjos87}.  The modeling of minimum bias
events  in this picture assumes that in a soft hadron reaction the
incoming

particles exchange color. After interacting, each hadron has broken
into
two remnants, a quark and a diquark. Two color strings stretch  from

one forward to one backward remnant. In a proton-proton reaction,
each string connects a quark to a diquark

forming a color singlet.  When these strings fragment, and form
particles,
 they produce an approximately flat rapidity spectrum of mostly
mesons, and a pair of leading baryons which carry the original
diquarks
from the ends of the string.   At high energies minijet production

changes this simple picture.  Typical high energy events contain
multiple

interactions between the partons\cite{heur85,dura,bloc90}.
Often these interactions are hard enough
to add significant transverse momentum to the event, and can be
treated

perturbatively. These are the minijets.  In string models minijets

are added to the basic 2-string configuration to account for the
increase
of average transverse momentum, the $p_t$-multiplicity correlation,

and the excess of events at high
multiplicity\cite{gais89,wang91,flet91b}.

In typical high  $p_t$ events the  underlying event  is treated in a
similar
way.  For instance, when a two jet system is formed,  the beam
particles

`lose' a parton to the 2-jet system, so the beam fragments carry
color.
 In order to form a color singlet the forward and backward beam
fragments
must be  joined by a string.  Again, this fills the central rapidity

region with particles.  Minijets appear again in the underlying
event.
Secondary interactions between partons can produce minijets
underlying

the main, high $p_t$ reaction.  Finally, initial state radiation and
final
state showers add more particles to the event.

W-boson fusion events have a different structure. In these events
the incoming quarks radiate a color singlet W-boson, so the recoiling
quark
which  forms a forward jet, and the surviving beam remnant form a
color
singlet.
Since these partons are both in the forward (backward) region, they
won't
create particles in the central region.    A simple example of this
structure is shown in figure 2. The figure shows a plot in
pseudorapidity,
$\eta$
and azimuthal angle, $\phi$, of the particles produce from  the
fragmentation
of two color strings. Each string stretches between a 40 GeV $p_t$
quark jet at
pseudorapidity of 4 and a beam remnant moving along the z axis. The
plot
shows ten SSC  events, each with the same parton level configuration,
but
fragmented separately using the JETSET Monte-Carlo\cite{sjos87b}.
 The gap is clearly
evident.   Higgs  production events will have this structure if the

decay products of the Higgs can be removed.
This configuration is also expected from jet production

via the exchange of a neutral boson, (W, or Z) or
pomeron\cite{cheh92}.

When two quarks interact to produce a  Higgs by $W$ fusion,
the remaining clouds of quarks and gluons that form the protons must
pass
through one another. If there are secondary interactions, either soft
color
exchange  or minijet production,  the gap will be destroyed.
 This leads to a small survival  probability for  gaps.

 The minijet model provides a probabilistic  interpretation
for  parton \linebreak[4]
exchanges\cite{dura,bloc90,gais89}.
When two protons pass through one another at an impact parameter $b$,

the average number of interactions

is given by

\begin{equation}
n (b)  =   \sigma_0 \  A(b,\mu_1) + \sigma_{jet} \  A(b,\mu_2)

\end{equation}
$\sigma_{jet}$ is the cross section for producing hard jets at scales
greater than a cutoff $t_{min}={\rm 5.3 GeV^2}$, calculated using the
proton
structure functions of ref{\cite{duke84}.
  $\sigma_0={\rm 125 GeV^{-2}}$
 is a constant  representing softer exchanges, chosen to
reproduce the low energy total cross section. $A(b)$ is the overlap

function of the  two protons in impact parameter space given by

\begin{equation}
 A(b,\mu) = \frac{\mu^2}{12\pi} \frac{1}{8} \left( \mu\,b\right)^3\
K_3
         (\mu\,b)  \ \ \mu_1= .85 \ \mu_2=.76
\end{equation}
 Assuming that the
interactions are independent, the distribution in the number
of interactions is poissonian, so the probability of having no
interactions

for impact parameter b is given by  ${\rm exp[-n(b)]}$. In this model
the inelastic
and total cross sections are given by\cite{dura},
\begin{eqnarray}
\sigma_{inel} & =& \int d^2 b \ \  1-e^{-n(b)}   \\
\sigma_{tot}  & =& 2 \int d^2 b \ \  1-e^{-n(b)/2}
\end{eqnarray}
If we assume that the distribution of $W$'s in impact parameter space
in the

proton is the same as for the partons which produce minijets, the
inclusive

differential cross
section in impact parameter space for Higgs production through W
fusion is just
\begin{equation}
\frac{d\sigma}{d^2b} = \sigma_{WW->H} \  A(b,\mu_2)
\end{equation}
and the cross section for producing a Higgs {\em without} having any
secondary
interactions is

\begin{equation}
\frac{d\sigma_{gap}}{d^2b} = \sigma_{WW->H} \ A(b,\mu_2) \times
e^{-n(b)}
\end{equation}
The exponential factor suppresses the contribution to the cross
section
where the two protons overlap and there  is likely to be a secondary

collision.  The integral of  eq. (6), divided by the total production
cross
section gives a  survival probability for the gap.  Using these
expressions
 we  can calculate the total cross section,
elastic slope, and survival probability at different energies.
The  two values of $\mu$ allow the partons contributing to the

soft and hard parts of the cross section to have slightly different

impact parameter distributions. The  value of $\mu_2$, and $t_{min}$
are  chosen to fit the values of the total cross section and
elastic slope measured at the Tevatron\cite{amos90}:

$\sigma_{tot}=72.1\pm 3.3$
and $B_{el}= 16.5 \pm .5$.
The results are given in Table 1.

 The last column gives the survival
probability  under the assumption that the distribution of W's in the

proton follows the soft form factor. This leads to a  smaller
survival

probability.

\begin{table}[P]
\centering
\caption{Minijet model predictions for scattering parameters and
survival

probability.}
\begin{tabular}{c|ccc}
  $E_{cm}$ &{$\sigma_{total}$}&{Elastic Slope, $B$}& Survival
Probability\\

   $ [GeV]$ &    [mb]        &  $ [GeV^{-2}]$       &     \\ \hline
   63.0   & 42.   &  13.3  & 14.\%   \\

  630.    & 58.   &  14.9  & 10.\%  \\

  1800.   & 72.   &  16.0  & 7.8\%   \\

  40000.  & 137.  &  21.0  & 3.3\%  \\

 \end{tabular}
  \label{survival}
\end{table}

In Pythia the effect of multiple interactions on the event is
explicitly

included by producing secondary, low $p_t$, minijet pairs. We have

run separate samples of events with and without multiple
interactions.
  If multiparton interactions in the PYTHIA model\cite{sjos87a}
 are turned on, 96 \% of events have a secondary interaction which

changes the color flow in the event,{\em i.e.} PYTHIA predicts

a survival probability of 4 \%, in agreement with our estimate and
that
of reference \cite{bj92}.

\section{Gap production and radiation}
We have seen that when there is a color singlet exchange

between two interacting protons, and there are no secondary
interactions,

particle production from the underlying event, beam fragmentation,
is
suppressed, causing a gap. Equally important is the fact that

parton, generally gluon,  radiation in the  region between the

tagging jets is also suppressed.

The foundation of all rapidity gap physics is that the structure of
an
event is dependent on the color structure of the interaction.  Since
this dependence is so crucial, it is important to understand the
physics involved, how this physics in manifest in a perturbative
calculation using Feynman diagrams, and how it is implemented in
Monte-Carlo event simulations.  A simple example will be useful in
investigating these facets of gap physics. We will

consider the 2$\rightarrow$2 process u+c $\rightarrow$ u+c.  There
are two possible $t$ channel exchanges between the incoming partons;
gluon

exchange (QCD) and photon exchange (QED). The feynman diagrams are
shown in

figure 3.

The difference in the radiation in QED and QCD exchange events can

be understood from a simple classical analogy.
{}From electrodynamics, we know that accelerated charges radiate, and
tend to radiate collinearly.  In QCD

when a color charge is accelerated, it will result in
radiation, primarily tangentially to the path of the accelerated
charge.
There are 2 relevant angles  in the scattering, $\theta_{scatter}$

the angle through which the fermion scatters, and

 $\theta_{charge.}$ the angle through which the color charge

scatters.   In single gluon or photon exchange, the cross section

is large when $\theta_{scatter}$ is small, ie small $\hat{t}$.

In single photon exchange, $\theta_{scatter}=\theta_{charge}$,
 the color charge is typically accelerated through

angle $\theta_{scatter}$ from the initial
direction, and we expect gluon radiation
only in the regions collinear to this acceleration $i.e.$ we only
expect

radiation when

$ \theta_{radiation} \approx \theta_{scatter} $. Hence there will be
a
large  region of rapidity
in which radiation is
suppressed.

  However for QCD exchange, the {\em charge} is accelerated from
the incoming up quark direction to the outgoing charm quark direction
or $\theta_{charge} = 2\pi - \theta_{scatter}$.
If we follow the classical trajectory of the charge (not fermion), it

would be a continuous curve connecting the direction of the incoming

up quark to the outgoing charm quark.
For small $\theta_{scatter}$ the tangents to  this curve
cover essentially the entire angular region,
so radiation can be produced over the entire rapidity range.

We  can see this effect in perturbation theory
by calculating  the $2 \rightarrow 3$ matrix elements\cite{elli87},
and

plotting the ratio of the
QED to QCD contributions for fixed quark jets, while varying the
rapidity of the radiated gluon. The feynman graphs for the two
processes are given by adding an external gluon line
to the diagrams in Fig. 3.
Figure ~4 shows the resulting suppression of radiation in the gap
from
color singlet (photon) exchange compared with gluon exchange.
We fix the final quarks at pseudorapidity, $\eta= \pm 5$
and $p_t=40 GeV$.  The gluon's $p_t$ was fixed at 1 GeV and its
rapidity
varied from -10 to 10. All three final

particles where set in a plane.  The quark leg which was furthest
from

the final state gluon was

given additional $p_t$ to balance the gluon, and

the remaining degrees of freedom where
used to conserve energy and momentum.

Near the directions of the
outgoing fermions, the matrix elements are  dominated
by initial and final state radiation terms, and the

photon exchange, and gluon exchange matrix elements are similar in

shape. This is clear from the excellent cancelation of the
singularities
in the matrix elements at $\eta_{radiation}=\eta_{jet}=5$ in Fig. 4.

When the radiated gluon is in the central
region, far from the outgoing fermions, destructive interference in
the photon exchange matrix element

suppresses radiation into the gap relative to the
gluon exchange process. At high rapidities, beyond the
rapidity of the  quark jets, destructive interference
in the QCD exchange causes a bump in the ratio.
The above analysis was also performed for gluon

energies of 10 and 20 GeV and the graph has the same properties.

The difference in the radiation in these two processes comes from the
 different interference between initial and final state radiation

graphs, and by radiation off the t-channel gluon propagator.

In general there
is interference between graphs where the gluon is radiated

from the same color line. In  photon exchange  this means
that the interference terms come from gluon radiation  off a single
fermion line. In the gluon exchange case the interference is
between radiation off different fermion lines, one initial

one final, or between radiation of a fermion line, and  the gluon
propagator.  These interference terms  cause the different
radiation patterns.

There is also an interference term between the photon exchange and
gluon exchange diagrams when a  gluon is

radiated off of  different fermion lines.  This adds a small
correction, order $\alpha \alpha_s$, to  the non-gap forming cross
section,
which is not included in the figure.

Monte-Carlos require a quick method for determining the probability
for radiation.  They approximate the above interference by using
angular ordering.  The angular ordering algorithm places a limit on
the
phase space into which  partons can radiate. A cone is drawn around

each parton line, whose angle is determined by the angle between that
parton,  and the parton  which is color connected to it
in the $1/N_c$ approximation.  Radiation in

the cone is allowed in the normal leading log approximation.
Radiation
is not allowed outside the cone.

In the production of large gaps, in the small angle approximation,
 this constraint has a simple
form. Radiation is allowed in a cone of half angle $\theta_{charge}$
(see Fig. 3)
around the quark jet, or up  to an angle $2\theta_{charge}$. In
terms of rapidity, radiation is approximately limited by

\begin{eqnarray}
\eta_{max} &= &ln \left[ Tan \left(\frac{2\theta_{color}}{2}\right)
\right]  \\
      & \simeq & ln \left[ \frac{\theta_{color}}{2}\right] -
ln\left[2\right]

      \simeq    \eta_{jet} - .7

\end{eqnarray}
The angular ordering algorithm replaces the
exponential falloff in the central region in Fig. 4 with a cutoff at
$\eta\approx 4.3$.

To summarize, radiation  is suppressed in the central region when
a color singlet  is exchanged between partons. This effect

can be understood classically, is present in the

$2\rightarrow 3$ feynman amplitudes, and is

predicted by the angular ordering algorithm

used by the HERWIG Monte-Carlo.

\section{Higgs Production in the HERWIG Monte-Carlo}
The  angular ordering algorithm used in the

HERWIG\cite{herwig,marc88}
Monte-Carlo  includes the color flow information needed

to correctly generate parton radiation in

gap events. Here we use HERWIG to study the

event structure in heavy higgs production in the

processes $WW\rightarrow h^0 \rightarrow WW$
and $gg\rightarrow h^0 \rightarrow WW$. We  compare these to

PYTHIA predictions for $gg\rightarrow h^0$, $q\bar{q}\rightarrow WW$
and $gg\rightarrow t\bar{t} \rightarrow WW$. In each case,

after generating an event, the W's are eliminated

from the event record and excluded from the remaining analysis.
We choose $m_h^0= 500 GeV$ and $m_t=150 GeV$.
For the background processes, direct W pair production and top decay
to W's,
we require the W-pair mass to be between 450 and 500 GeV. We have

not done a detailed  simulation of  the reconstruction of the

W's or of other detector acceptances.

Here we do a simple analysis to show the effect
of the gap requirement.
For  each event we count the total multiplicity in the central

rapidity range, $\eta< 2$,
(charged and neutral $\pi's, k's, \eta's$).
Figure 5 shows the HERWIG multiplicity distribution

for WW fusion to higgs, and $gg \rightarrow h$, and
Pythia predictions for the backgrounds from

direct $W$ pair production and  top pair decay to W's.

  The signal from W fusion to the Higgs   is a peak at

 multiplicity of zero. 78 \% of the signal in is  the region

of multiplicity less
than  or equal to two.

 The curves are normalized to

the cross section given by HERWIG or PYTHIA, multiplied by the

 branching ratio squared for the decay of that state to

$W+W- \rightarrow l^+ \nu l^- \bar{\nu}$, and for gap production,

by the survival probability of 3\%.  For $WW$ fusion to Higgs
this leaves a cross section of .72 fb.   The total rate for direct
$W$ pair production  with leptonic decays is 340 fb, but only .01\%
of these have  a gap, so the background is small.
Top decays to $W$'s have a large cross section, and calculating

the small fluctuations  directly is difficult. The

curve in Fig. 5 is a negative binomial fit to  the multiplicity

distribution from
24000 events, where the $W$ pair is within 50 GeV of the Higgs mass.
The requirement that there be fewer than 2 particles in the central

region  supresses the top quark contribution by six
orders of magnitude, making the background small.

A more detailed analysis using
jet tagging  to define  the gap  should  improve the background
rejection, and might help improve the signal by eliminating the long
tail on the
$WW\rightarrow H$ multiplicity distribution. However, it also leads
to the rejection of some signal events whose tagging jets are outside
the

range of detectors.  Jet tagging, jet tagging with a gap requirement,
and a simple study of the central multiplicity should all be
complimentary

in an eventual Higgs search.

\section{Conclusions}

Events  containing  a $W$ boson pair generated by

boson-boson fusion at the Higgs resonance at the SSC will have a
completely different underlying event structure than other $W$ boson
pair events. This difference is caused by the properties of QCD
radiation
in  color singlet exchange  processes and by the lack of a soft  soft
particle production in the central region.

Many events of this type will not ``survive" because of

spectator parton interactions.

We have calculated  this survival probability

in the minijet model. The survival probability

drops at high energies because of the rise of the total and inelastic

cross sections. At the SSC about 3\% of  Higgs production events

have no secondary interactions.

In one year at the SSC we expect about 5 events containing a 500 GeV

Higgs decaying through $W's$  to the

``silver plated" mode, electrons, muons, and neutrinos.

Requiring that there be a true gap, containing few, or

no  particles in  the central region of the event  supresses
the background from top decays and direct $W$ pair production
by  three to six  orders of magnitude.  Essentially, there is no

background from these processes.  It is not necessary to do
jet tagging in this case to determine the true edges of the gap.

There are still unresolved problems in making gap production a

realistic signal  at the SSC.  First, experiments
must be able to measure

soft particle production reliably. Events with as few as 10 soft
particles
need to be excluded to keep them from overwhelming the gap signal.
Second, the experiments must deal

with event overlap.   A minimum bias event overlapping the Higgs  gap
event
will fill the gap with particles.

Unless the tracking is good enough to reliably assign

soft particles to the correct vertex, studying gap physics will
only  be possible   at low luminosities when event overlap is small.
But the low rate for higgs production makes this approach impossible
unless the total event rate can be increased.  This requires that
we be able to reconstruct  the hadronic decays of the $W$'s,
or that  our model for the survival probability is wrong.
Either of these eventualities might increase the

signal rate by an order of magnitude.

\newpage
\section*{Acknowledgments}
We would like to thank Francis Halzen, Dieter Zeppenfeld, B.J.
Bjorken,
Tjorbjorn Sj\"ostrand, V.A. Khoze
and Alan Stange for useful comments.

This research was supported in part  by the U.~S.~Department of
Energy
 under contract No.~DE-AC02-76ER00881.
\newpage

\clearpage
{\bf Figure Captions}

Figure 1.  Higgs boson production by gluon fusion (a), and by weak
boson

fusion (b).

Figure 2. "Lego-plot" for ten  superimposed two jet events,
 with a central gap, modeled
with string fragmentation.

Figure 3.  Graphs for single gluon exchange, and single photon
exchange,  labeled with color and fermion scattering angles.

Figure 4.  Ratio of photon vs. gluon exchange matrix elements
for $qq' \rightarrow qq`g$, as a function of the pseudorapidity
of the gluon, for quarks fixed at $\eta = \pm5$,
showing the supression of gluon radiation in
color singlet exchange.

Figure 5.   Multiplicity distributions  of particles inside $\eta=2$

for Higgs production and
background processes. Curves are normalized
to the cross section times branching ratio to  pure leptonic decays
of the W.

\end{document}